\documentclass[11pt,a4paper]{article}
\usepackage[pdftex]{graphicx}
\usepackage{latexsym}
\usepackage{url}
\usepackage{here}
\usepackage{mathrsfs}

\begin{document}

\title{Search for developments of a box having multiple ways of folding by SAT solver}

\author{Riona Tadaki\thanks{Department of Computer Science, Gunma Univ., Tenjin 1-5-1, Kiryu, Gunma 376-8515, Japan} \thanks{t181d038@gunma-u.ac.jp} \and Kazuyuki Amano\footnotemark[1] \thanks{amano@gunma-u.ac.jp}}

\maketitle

\begin{abstract}
	A polyomino is called a development if it can make a box by folding edges of unit squares forming the polyomino.
	It is known that there are developments that can fold into a box (or boxes) in multiple ways.
	In this work, we conducted a computer search for finding such developments by using a SAT solver.
	As a result, we found thousands of such developments including a polyomino of area 52 that can fold into a box of size $1 \times 2 \times 8$
  in five different ways.
\end{abstract}

\section{Introduction}

A {\it polyomino} is a two-dimensional shape formed by joining unit squares edge to edge. 
A polyomino is called a {\it development} if it can make a box by folding edges of unit squares forming the polyomino. 

\begin{figure}[tb]
\begin{center}
\includegraphics[clip,scale=0.3]{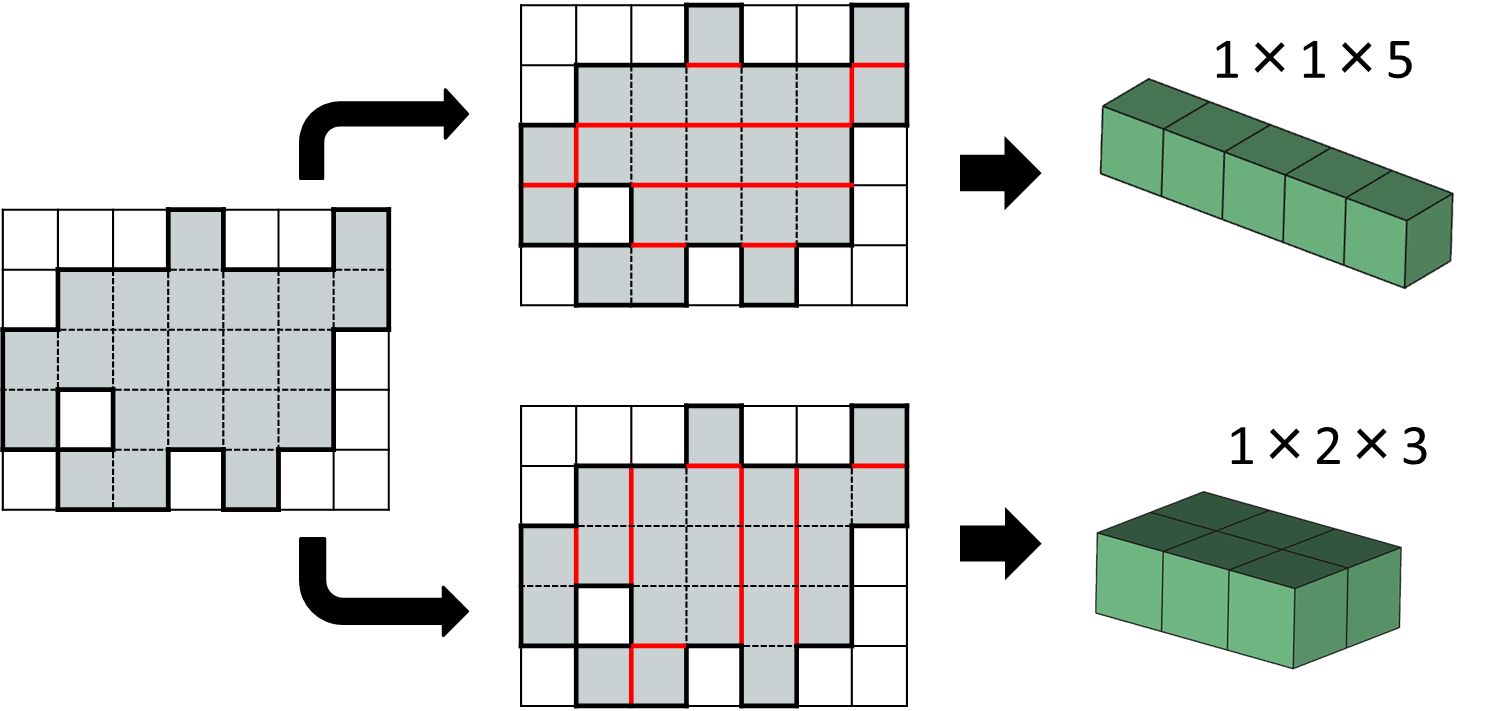}
\end{center}
\caption{A common development of two boxes.}
\label{fig:1-1}
\end{figure}

As in Fig. \ref{fig:1-1}, 
there are developments that can fold into two incongruent boxes. 
Many such developments have been discovered. 
For example, for the surface area 22, 
it was shown by an exhaustive computer search that
there are 2,263 common developments of two boxes 
of size $1\times 1 \times 5$ and $1\times 2\times 3$ \cite{Uehara2011}.

\begin{figure}[tb]
\begin{center}
\includegraphics[clip,scale=0.3]{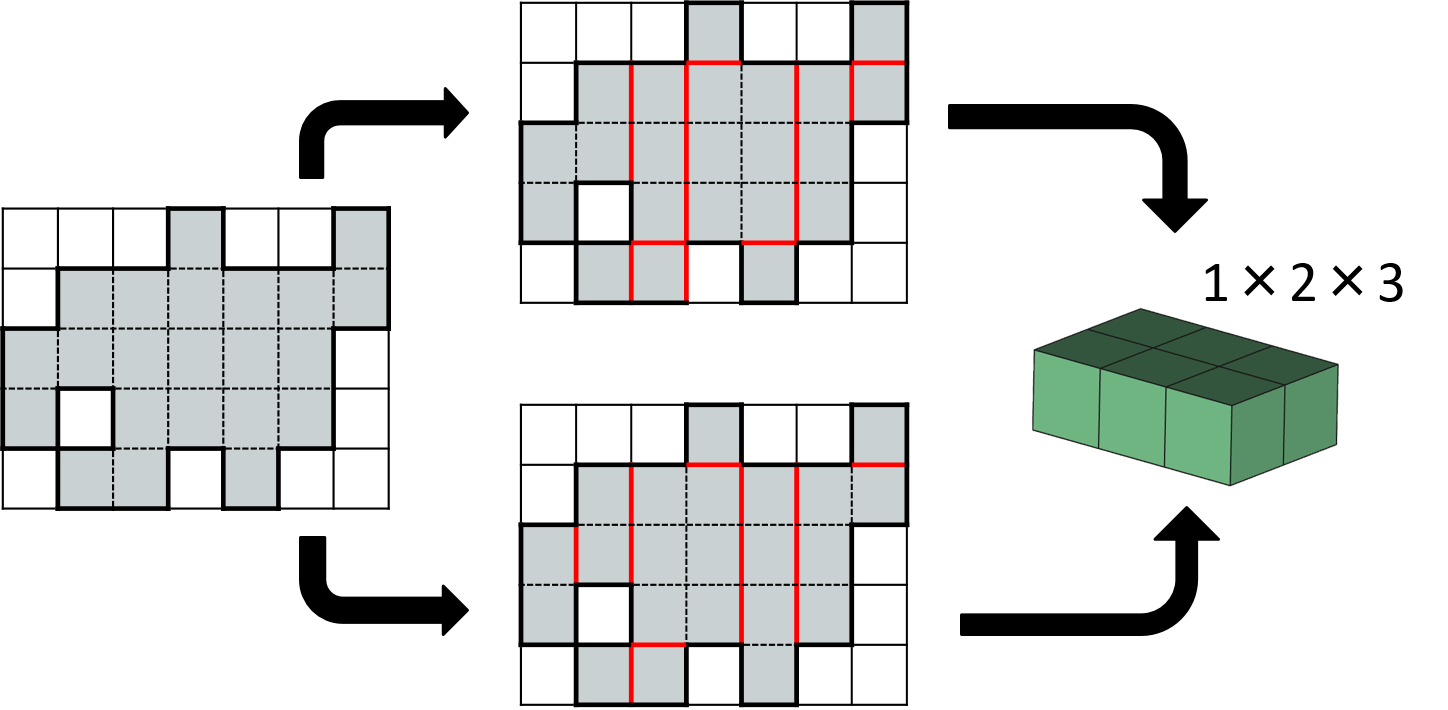}
\end{center}
\caption{A development of a box with two different foldings.}
\label{fig:1-2}
\end{figure}

\begin{figure}[H]
\begin{center}
\includegraphics[clip,scale=0.3]{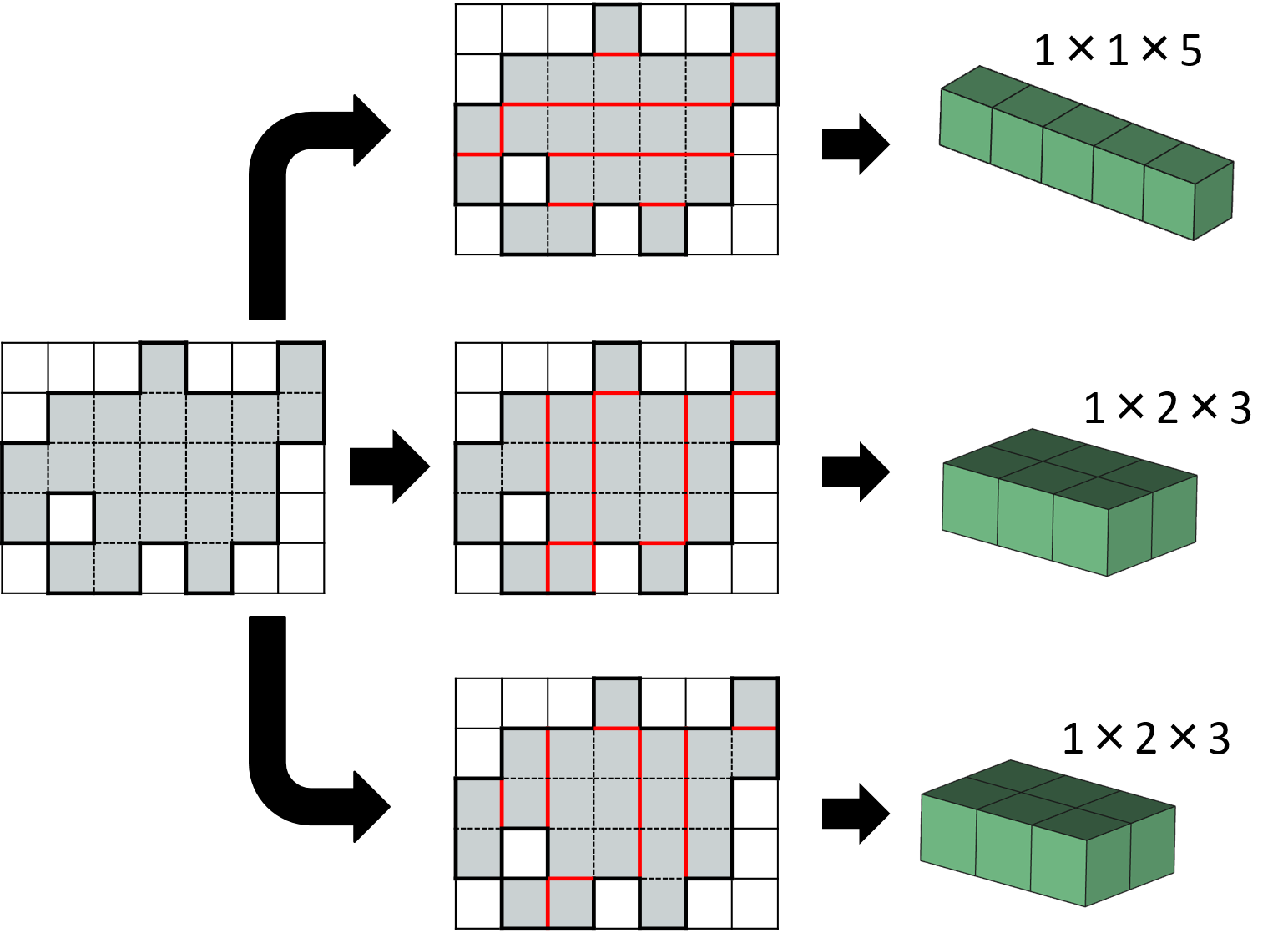}
\end{center}
\caption{A development of a box of with three foldings.}
\label{fig:1-3}
\end{figure}

There also are developments that can fold into a same box in multiple ways as shown in Fig. \ref{fig:1-2}. 
By noticing that the polyomino shown in Figs. \ref{fig:1-1} and \ref{fig:1-2} is fact identical, 
this development admits three different ways of folding into two boxes (see Fig. \ref{fig:1-3}).
It is also known that a polyomino of area 532 that can fold into three different boxes of size $2 \times 13 \times 58$, $7 \times 14 \times 38$ and
$7 \times 8 \times 56$ \cite{SU13}.
To the best of the authors' knowledge,
it is an open problem to see  whether there exists a common development of four (or more) different boxes.
See also, e.g., \cite{DO07, Uehara2019, Ueh14, XH17} and the references therein for other results on this fun topic.

In this work, we only consider {\it orthogonal} foldings, i.e.,
foldings such that all creases are on the edges of unit squares forming a polyomino.
We conducted a computer search for finding developments that can fold into a box (or boxes) in many ways
by using a SAT solver.
As a result, we found thousands of such developments including a polyomino of area 52 that can fold into a box of size $1 \times 2 \times 8$ in five
different ways (see Fig. \ref{52_5}).

In Section 2, we give a brief explanation of SAT solvers.
In Section 3, we show the SAT encoding of the problem of finding a polyomino that folds into a box (or boxes) in multiple ways.
The experimental results are shown in Section 4.
Finally, we close this report with a brief discussion in Section 5.

\section{SAT solver}
A SAT solver is a software that determines the satisfiability of a given Boolean formula, which is usually
given in Conjunctive Normal Form (CNF).
The CNF formula is a conjunction of clauses, where a clause is a disjunction of literals, where a literal is a variable or its negation.
See (\ref {eq:0}) for example.

\begin{equation}
(x_1\vee x_2 \vee x_3 \vee x_4) \wedge (x_1\vee\overline{x_3}) \wedge (\overline{x_2} \vee x_3)
\label{eq:0}
\end{equation}

If an input formula is satisfiable (SAT), a SAT solver outputs one satisfying assignment of the formula; 
if it is unsatisfiable (UNSAT), it says ``UNSAT''.
The DIMACS format is used as a standard input format. 
In this format, the CNF formula (\ref{eq:0}) is represented as follows.
\begin{table}[htb]
\begin{center}
\begin{tabular}{l}
{\tt p cnf 4 3} \\
{\tt 1 2 3 4 0 }\\
{\tt1 -3 0 }\\
{\tt-2 3 0 }\\
\end{tabular}
\end{center}
\end{table}

The first line describes the number of variables and the number of clauses as a header.
Given this CNF to a SAT solver, it outputs an assignment that makes the formula true like the following.
\begin{eqnarray*}
\begin{array}{l}
{\tt SAT}\\
\mbox{{\tt 1 -2 3 -4 0}}
\end{array}
\end{eqnarray*}


This says that the assignment $ (x_1, x_2, x_3, x_4) = (1,0,1,0) $ satisfies the CNF formula (\ref {eq:0}).
\section{SAT Formulation}
In this section, we show how to encode the problem of finding a common development of boxes into a SAT problem.

\subsection{Variables}
\label{sec:var}

Suppose that we are aiming to find a development of
an $a \times b \times c $ box placed on an $n \times n$ board.
We introduce three types of Boolean variables as follows.

\begin{figure}[tb]
\begin{center}
\includegraphics[clip,scale=0.6]{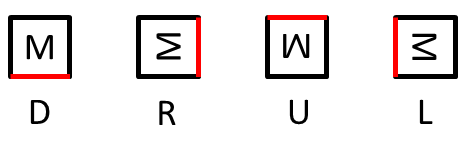}
\caption{Direction of putting a unit square of the box into a cell of the board.}

\label{fig:3-1}
\end{center}
\end{figure}

\begin{description}
	\item[Type 1:]
	For each unit square of the board, we assign a Boolean variable $i$.
	The variable $i$ is true if the corresponding unit square is occupied
	by a development, and is false otherwise.
	The number of variables of this type is $n^2$, i.e., the area of the board.
	
	\item[Type 2:]
	For each unit square $m$ of the box, for each direction
	$r \in \{D,R,U,L\}$ and for each cell $i$ of the board, we assign
	a Boolean variable $mi_r$ indicating whether the unit square $m$ of the box is
	mapped to the cell $i$ of the board with direction $r$ (see Fig. \ref{fig:3-1}).
	The number of variables of this type is $4n^2f$, where $f$ is the surface area of the box that is equal to $2(ab+ac+bc)$.

	\item[Type 3:]
	For each edge of unit squares of the box, we assign a Boolean variable $e$
	indicating whether the edge corresponding to $e$ is cut in a development.
	The variable $e$ is true if the corresponding edge is cut,
	and is false otherwise.
	The number variables of this type is $2(ab+ac+bc+a+b+c)$.
\end{description}
Note that the variables of Type 2 are dominant.

\subsection{Clauses}
We need several sets of constraints to express
the problem.
Let $M$ be the set of all unit squares of a box, $I$
be the set of all cells of the board, and 
$\mathscr{R}$ be the set of four directions, i.e.,
$\mathscr{R} := \{U,R,D,L\}$.

\subsubsection{Conditions for adjacent unit squares and edges}

\begin{figure}[tb]
\begin{center}
\includegraphics[clip,scale=0.5]{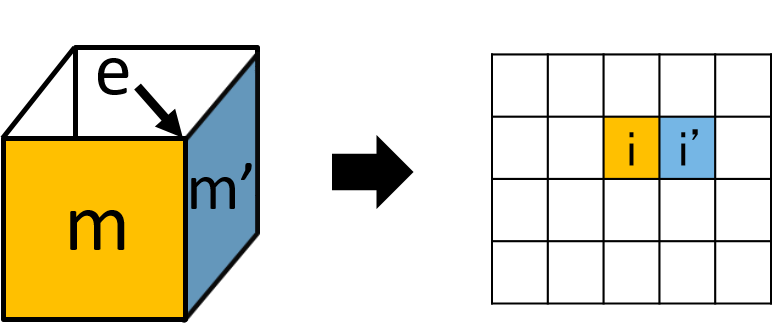}
\caption{The mapping of adjacent unit squares to cells of the board.}
\label{fig:3-3}
\end{center}
\end{figure}

Suppose that a variable $mi_r$ is true for some $m \in M$, $i \in I$ and
$r \in \mathscr{R}$, which means
that a unit square $m$ of a box is mapped to a cell $i$ in a board with direction $r$.
Suppose also that the unit square $m$ is adjacent
to a unit square $m' \in M$ through an edge $e$ as in Fig. \ref{fig:3-3} (left).

If the edge $e$ is {\it not} cut in a development, then the unit square $m'$ is
mapped to an adequate cell $i'\in I$ with an adequate direction $r' \in \mathscr{R}$. 
See Fig. \ref{fig:3-3}.
We can express this as the following set of constraints.


\begin{eqnarray}
\label{eq:3-2}
\forall_{m \in M}\,\forall_{i \in I}\,\forall_{r \in \mathscr{R}}\,, mi_r \wedge \bar{e} & \rightarrow & m'i'_{r'},
\end{eqnarray}
Note that, given $m \in M$, $i \in I$, $r \in \mathscr{R}$ and an edge $e$, each of $m' \in M$, $i' \in I$ and $r' \in \mathscr{R}$ are 
uniquely specified in a natural way.
We can express each constraint in (\ref{eq:3-2}) by a clause as
$(x \wedge y \rightarrow z) = (\overline{x} \vee \overline{y} \vee z)$. 

We also add the following set of constraints.
\begin{eqnarray}
\label{eq:3-3}
\forall_{m \in M}\,\forall_{i \in I}\,\forall_{r \in \mathscr{R}}\,, mi_r \wedge \bar{e} & \rightarrow & i'.
\end{eqnarray}

Note that the constraints (3) are, in fact, redundant as the constraints (2) and
(\ref{eq:3-6}) (which will be described later) can induce them. However, we place them in order to help SAT solvers to find a solution.

Next, we show the constraints for the case that the edge $e$ is cut in a development.
In order to avoid that separated unit squares are
mapped to adjacent cells, we add the following set of constraints.

\begin{equation}
\forall_{m \in M}\,\forall_{i \in I}\,\forall_{r \in \mathscr{R}}\,, mi_r \wedge e \rightarrow \bar{i'}
\label{eq:3-1}
\end{equation}

The constraints (\ref{eq:3-1}) express that
when we map a unit square $m \in M$ to a cell $i \in I$ with
direction $r \in \mathscr{R}$ and cut an edge $e$, 
we should not use an adjacent cell $i' \in I$ where $i'$ is uniquely specified when $m \in M$, $i \in I$, 
$r \in \mathscr{R}$ and the edge $e$ are given.

\subsection{Constraints of board}
The followings are the constraints to add to form a correct development on the board.
\begin{eqnarray}
\label{eq:3-4}
\forall_{m\in M}, \sum_{i \in I,r \in \mathscr{R}} mi_r=1\\
\label{eq:3-5}
\forall_{i\in I}, \sum_{m\in M ,r \in \mathscr{R}} mi_r\leq 1\\
\label{eq:3-6}
\forall_{i\in I}, i \leftrightarrow \bigvee_{m\in M, r\in \mathscr{R}} mi_r
\end{eqnarray}

Constraints (\ref{eq:3-4}) guarantee that each unit square of the box
is mapped to exactly one cell on the board.
Constraints (\ref{eq:3-5}) guarantee that each cell of the board is mapped
by at most one unit square of the box.
Constraints (\ref{eq:3-6}) give the correspondence between the variables 
of Type 2 and of Type 3.

The following set of constraints prevents a development from being separated; we should guarantee that all the cells $i \in I$ 
that are set to be true are connected.
This can be implemented by introducing the notion of ``ignition time'' to each occupied cell on the board.

Pick a particular cell $i$ in the board, and ``ignite'' $i$ at time $t=0$. 
Then, for $t=1,2,\ldots$, each cell $i'$ is ignited at time $t$ if some of
the surrounding cells of $i'$ has ignited at or before time $t-1$.
If all the occupied cells are ignited after a
sufficiently large time, then these are connected.

Let $d$ be a parameter called {\it maximum distance} which will be determined in a later section.
For each $i \in I$ and $ t\in \{0,1,\dots ,d\} $, we introduce a variable
$i_t$ indicating whether the cell $i$ is ignited at or before time $t$.
For each $i \in I$, let $ S_i $ be the set of cells connected to the 
cell $ i $ including $ i $ itself.
The constraints for guaranteeing that all the occupied cells are connected can be written as follows.
\begin{eqnarray}
\nonumber
\forall_{i\in I} \forall_{t \in \{1,\ldots,d\}} , i_t & \leftrightarrow & i \wedge (\bigvee_{i'\in S_i} i'_{t-1})\\
\nonumber
\sum_{i\in I} i_{t=0} & = & 1\\
\nonumber
\forall_{i \in I}, i_{t=0} & \rightarrow & i
\end{eqnarray}

If there is only one box to be developed, this finishes the encoding.
By solving a SAT problem for a formula described above,
we can obtain a development of a box of a given size
on a board as a satisfying assignment to the formula.

When we are aiming to find a common development 
of two or more boxes (of not necessarily different sizes),
we need to put another set of constraints guaranteeing that
each development occupies the same set of cells on the board.
This is achieved by simply adding the following set of
constraints.
\begin{equation}
\forall_{i\in I}, i_1 \leftrightarrow i_2
\label{eq:3-10}
\end{equation}
Here, we name a variable $i$ for the first box as $i_1$ and the one for the second box as $i_2$, for each $i \in I$.

This can easily be generalized to the case
for three or more boxes.
If we want to search a common development of three boxes, we should add the following sets of constraints, where $i_3$ denotes a variable corresponding to the cell $i \in I$ for the third box.
\begin{eqnarray}
\nonumber
\forall_{i\in I}, i_1 \leftrightarrow i_3\\
\nonumber
\forall_{i\in I}, i_2 \leftrightarrow i_3
\end{eqnarray}

\subsection{Reducing the Variables}

Our model has a large number of variables even for small cases.
For example, for searching
developments of a box of surface area 22 on a $15 \times 15$ board,
we have $19,800$ variables of Type 2 alone.

\begin{figure}[tb]
\begin{center}
\includegraphics[clip,scale=0.35]{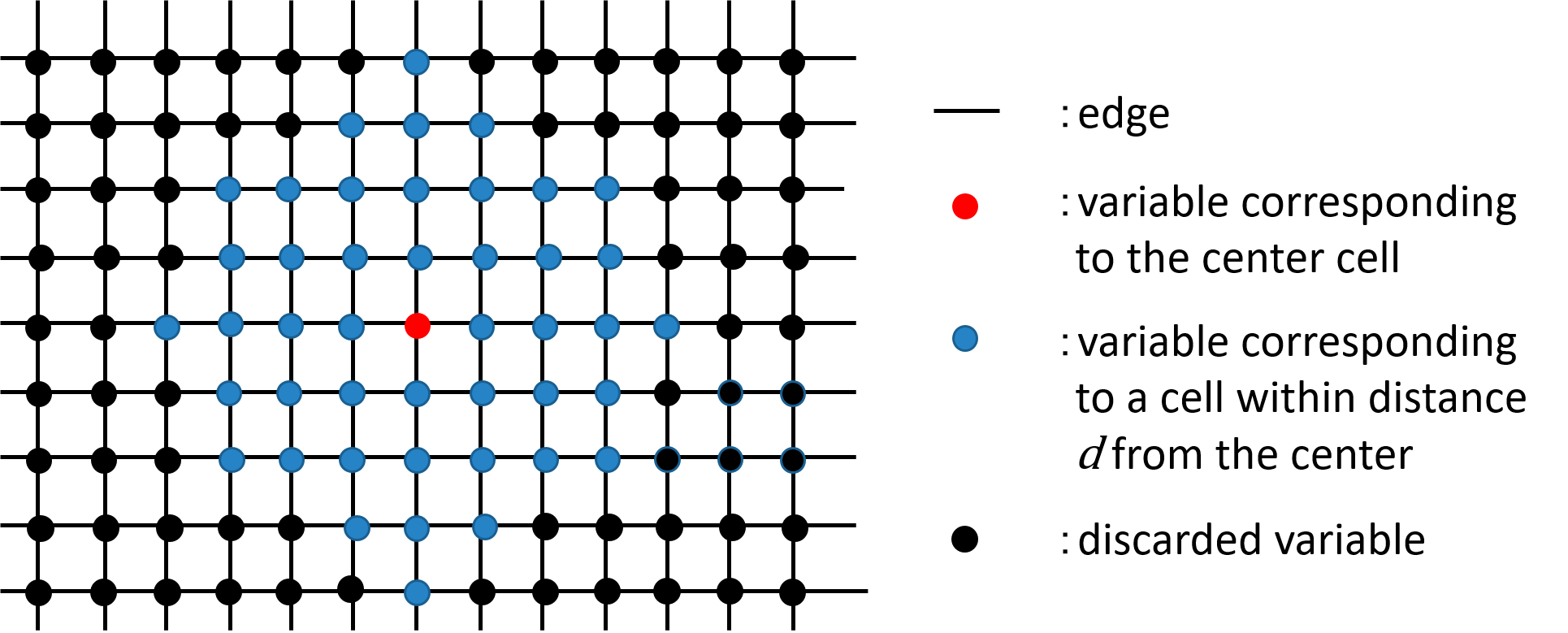}
\caption{Reduce the variables $mi_r$ by discarding
variables corresponding to cells located far from the center of the board.}
\label{fig:3-2}
\end{center}
\end{figure}

In order to speed up the search, we reduce the variables in the following way. 
Each variable of Type 2 is connecting to four adjacent variables of the same type, as the graph shown in Fig. \ref {fig:3-2}.
We pick one particular cell, which is located at the center of the board,
and then we discard all the variables corresponding to cells whose distance from the picked cell is larger than $d$, where the value of $d$
will be specified appropriately.
Note that if $d$ is set to be too small, then we will fail to find
a development although it actually exists. 

\clearpage
\section{Experiment}
The outline of our experiments is as follows:
We made a program that generates a CNF
formula in DIMACS format whose solution gives a common development of {\it two} (not necessarily different) 
boxes as explained in the previous section.
The choice of two here is made by some preliminary experiments; they suggest that solving a SAT problem
for finding a common development of three or more boxes seems quite time consuming.
Once we have obtained such a development, 
we apply an algorithm developed in 
\cite{Uehara2019} to
count the actual number of ways of folding a box.

We tried several good solvers in Parallel Track of SAT Competition 2017 \cite{SATCompetition} and used {\sf PLINGELING} developed by 
Biere \cite{Bie17} since it performs well for our models.
In order to speed up the search,
we generate and solve a number of CNF formulas obtained
by fixing some of the variables.
More precisely,
we pick a pair of unit squares from each of two
boxes and a pair of directions, and then try to find a
development such that two chosen squares are
mapped simultaneously to the center cell of the board with chosen directions.
We generate a CNF formula and run the SAT solver
for each choice of such pairs.
In the following experiments, we fix the size of a board to $15 \times 15$ and the maximum distance $d$ explained in the previous section to $15$.

\par
\par
\noindent

\begin{figure}[tb]
	\begin{center}
		\includegraphics[clip,width=6.24cm]{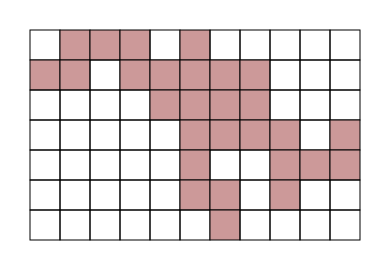}
		\caption{A development of a box of size $1 \times 2 \times 4$ having four different foldings. }
		\label{28_4}
	\end{center}
\end{figure}

\begin{figure}[tb]
	\begin{center}
		\includegraphics[clip,scale=0.35]{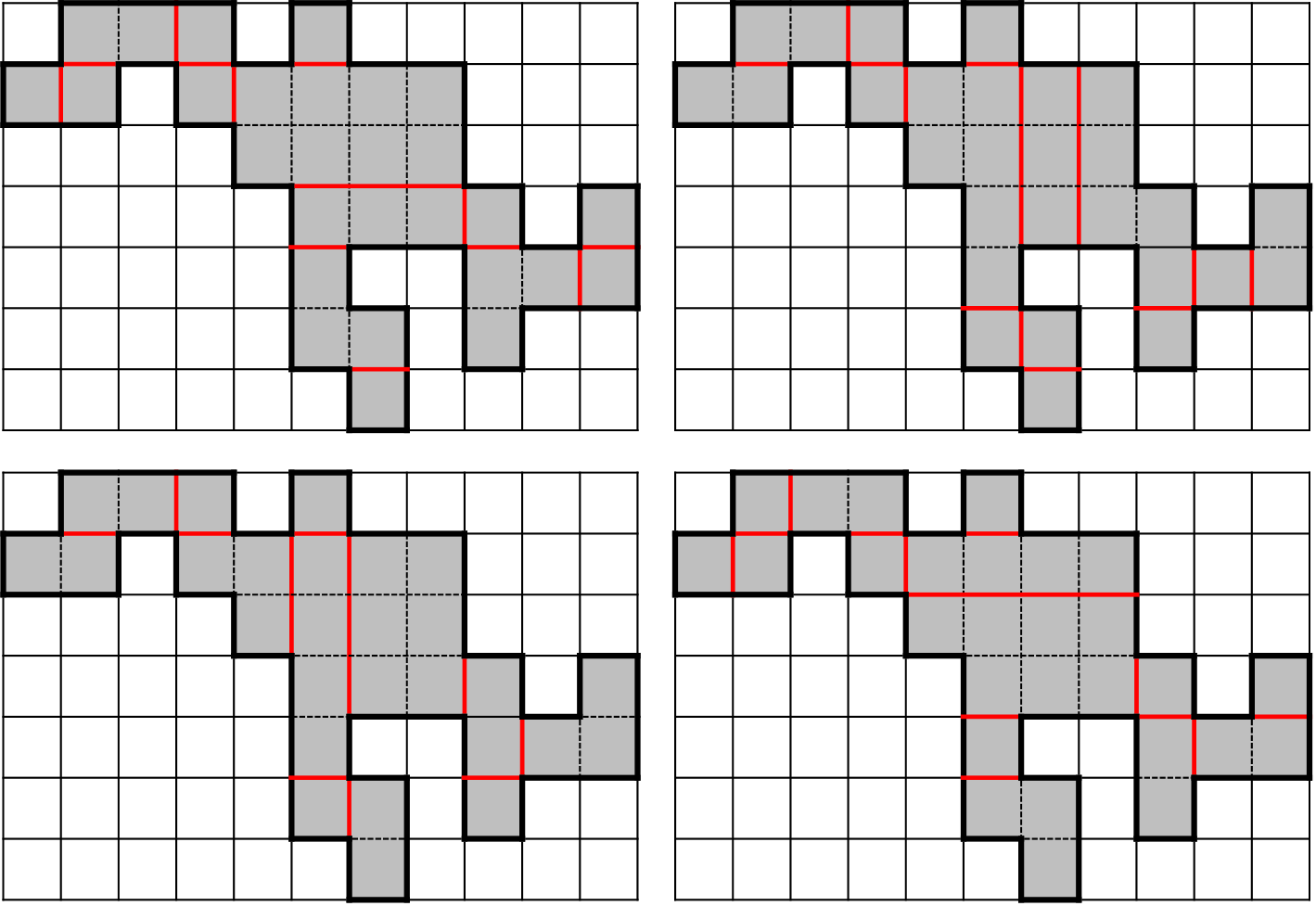}
		\caption{Folding lines of the development in Fig. \ref{28_4} to make a box of size $1\times 2\times 4$.}
		\label{28_4_2}
	\end{center}
\end{figure}

The first table, Table \ref{res28}, shows the results of 
experiments for searching
developments of area $28$ that can fold into a box of size $1 \times 2 \times 4$
in (at least) two different ways.
The numbers in the bracket in the second column of the table show
the numbers of non-isomorphic developments under
the rotation and reflection.
For example, the first line of the table says that
a SAT solver gives a development of a box of size $1 \times 2 \times
4$ 
having two different foldings 3376 times in our experiments, and out of which 2224 developments are non-isomorphic.
An example of a development that can fold into a box 
of size $1\times 2\times 4$ in four different ways is shown in Fig. \ref{28_4}.
The folding lines are shown in Fig. \ref{28_4_2}.

\begin{table}[tb] 
\caption{Result of surface area 28.} 
\label{res28}
\begin{center}
\begin{tabular}{ccc}\hline
size of box and number of foldings & quantity \\\hline\hline
	$1\times 2\times 4$ in 2 ways  & 3376 (2224) \\
	$1\times 2\times 4$ in 3 ways  & 428 (237) \\
	$1\times 2\times 4$ in 4 ways  & 11 (5)\\\hline
\end{tabular}
\end{center}
\end{table}

\begin{figure}[tb]
	\begin{center}
		\includegraphics[clip,width=6.24cm]{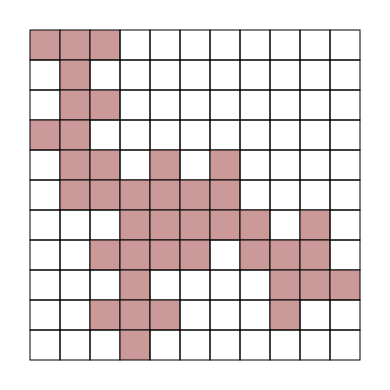}
		\caption{A development that can fold into a box of size $1 \times 2 \times 6$ in three ways and also a box of size $2 \times 2 \times 4$.}
		\label{40_1}
	\end{center}
\end{figure}

The next table, Table \ref{res40_1}, shows the results of experiments
for searching developments of area 40 that can fold into two boxes of
size $1\times 2\times 6$ and $2\times 2\times 4$.
A development that can fold into a box of size $1 \times 2 \times 6$ in three ways and also a box of size $2 \times 2 \times 4$
is shown in Fig. \ref{40_1}.
It may be fun to make boxes from this development.
Please try.

\begin{table}[tb] 
	\caption{Result of surface area 40 (part I).} 
	\label{res40_1}
	\begin{center}
		\begin{tabular}{cc}\hline
			size of boxes and number of foldings & quantity \\\hline\hline
			$1\times 2\times 6$ in 1 way $+$ $2\times 2\times 4$ in 1 way
			& 1949 (1458) \\
			$1\times 2\times 6$ in 2 ways $+$ $ 2\times 2\times 4$ in 1 way
			& 133 (107) \\
			$1\times 2\times 6$ in 3 ways $+$ $ 2\times 2\times 4$ in 1 way
			& 2 (1) \\\hline
		\end{tabular}
	\end{center}
\end{table}

\begin{figure}[tb]
	\begin{center}
		\includegraphics[clip,width=5.76cm]{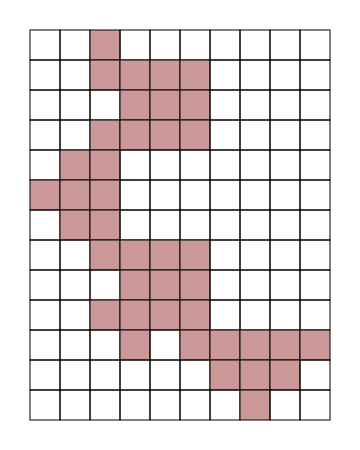}
		\caption{A development of a box of size $1\times 2\times 6$ having four different foldings. }
		\label{40_2}
	\end{center}
\end{figure}

Table \ref{res40_2} shows the results of experiments
for searching developments of area 40 that can fold into a box of
size $1\times 2\times 6$ in (at least) two ways.
An example of a development that can fold into a box of size $1 \times 2 \times 6$ in four ways is shown in Fig. \ref{40_2}.

\begin{table}[tb] 
	\caption{Result of surface area 40 (part II).} 
	\label{res40_2}
	\begin{center}
		\begin{tabular}{ccc}\hline
			size of box and number of foldings & quantity \\\hline\hline
			$1\times 2\times 6$ in 2 ways & 997 (767) \\
			$1\times 2\times 6$ in 3 ways & 104 (86) \\
			$1\times 2\times 6$ in 4 ways & 43 (36) \\\hline
		\end{tabular}
	\end{center}
\end{table}

\begin{figure}[tb]
	\begin{center}
		\includegraphics[clip,width=6.72cm]{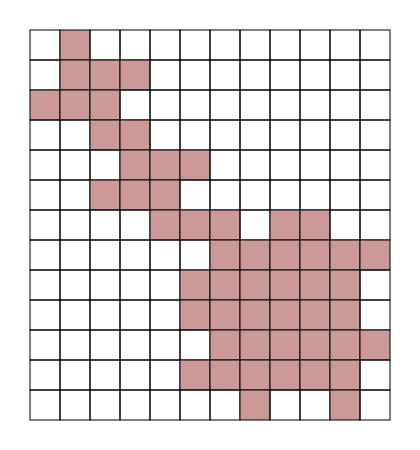}
		\caption{A development of a box of size $1\times 2\times 8$ having five different foldings. }
		\label{52_5}
	\end{center}
\end{figure}

\begin{figure}[tb]
	\begin{center}
		\includegraphics[clip,scale=0.5]{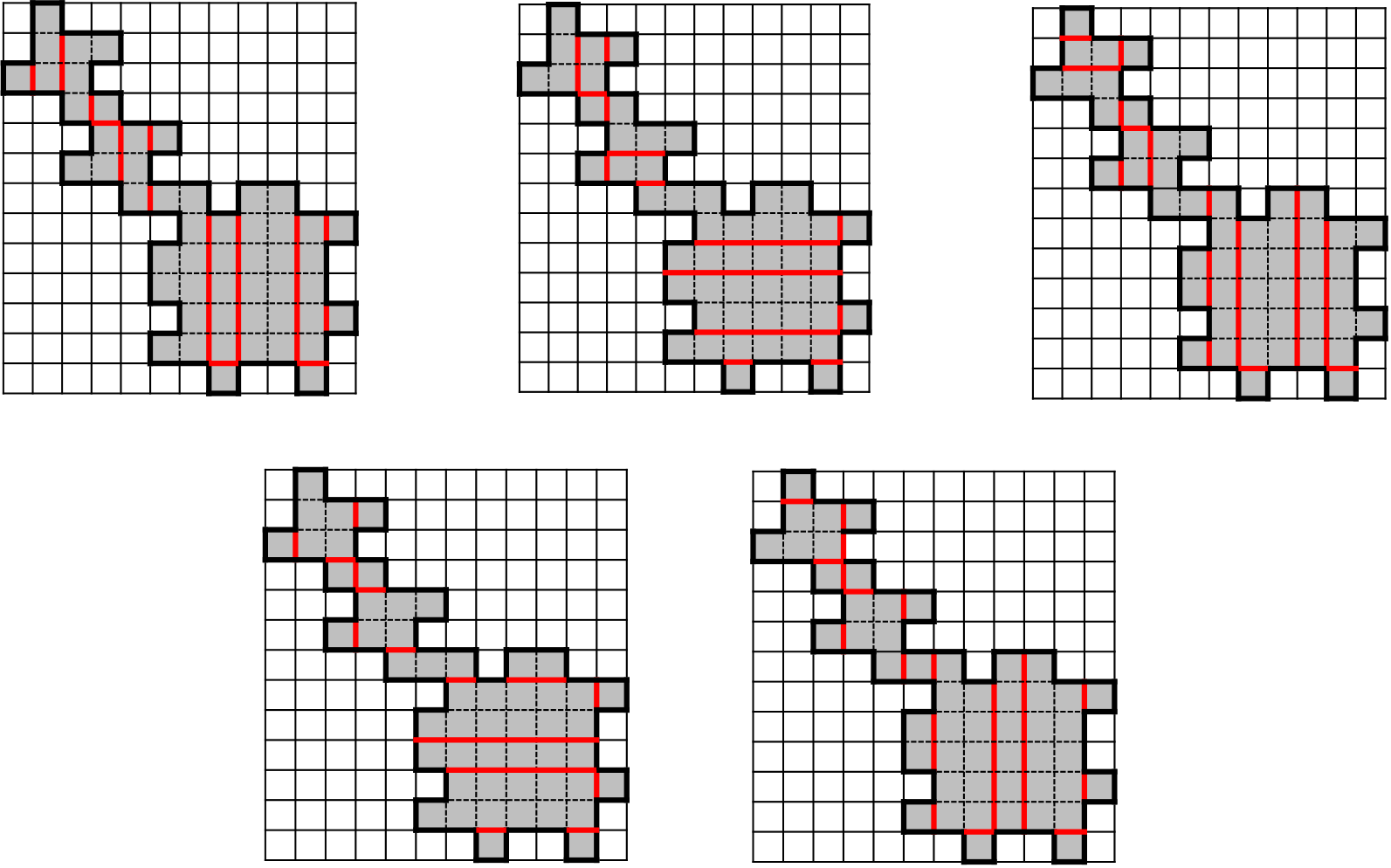}
		\caption{Folding lines of the development in Fig. \ref{52_5} to make a box of size $1\times 2\times 8$.}
		\label{52_5-2}
	\end{center}
\end{figure}

Table \ref{res52} shows the results of experiments
for searching developments of area 52 that can fold into a box of
size $1\times 2\times 8$ in (at least) two ways.
A development that can fold into a box of size $1 \times 2 \times 8$ in five ways is shown
in Fig. \ref{52_5} and its folding lines are shown in Figs. \ref{52_5-2}.
Note that this is the only development that has five ways of box folding obtained in our experiments.

\begin{table}[tb] 
	\caption{Result of surface area 52.} 
	\label{res52}
	\begin{center}
		\begin{tabular}{cc}\hline
			size of box and number of foldings & quantity \\\hline\hline
			$1\times 2\times 8$ in 2ways & 1122 (602) \\
			$1\times 2\times 8$ in 3ways & 744 (519) \\
			$1\times 2\times 8$ in 4ways & 40 (19) \\
			$1\times 2\times 8$ in 5ways & 8 (1) \\\hline
		\end{tabular}
	\end{center}
\end{table}

When the surface area is 52, a typical number of variables in a CNF formula is about 17,000, the number of clauses is about 1,000,000,  and the file size, which represents the size of a CNF formula,  is about 20MB.
The searching time for each development is varied; mostly between 200 and 10,000 seconds.

\begin{figure}[tb]
	\begin{center}
		\includegraphics[clip,width=7.68cm]{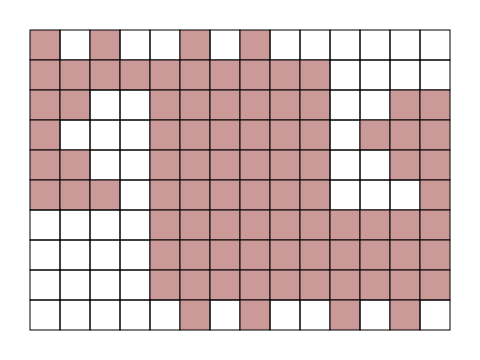}
		\caption{A development of a box of size $1 \times 4 \times 8$ having four different foldings.}
		\label{88}
	\end{center}
\end{figure}

Finally, Table \ref{res88} shows the results of experiments
for searching developments of area 88 that can fold into a box of
size $1\times 4\times 8$ in (at least) two ways.
An example of a development that can fold into a box of size $1 \times 4 \times 8$ in four ways is shown in Fig. \ref{88}.

\begin{table}[tb] 
	\caption{Result of surface area 88.} 
	\label{res88}
	\begin{center}
		\begin{tabular}{cc}\hline
			size of box and number of foldings & quantity \\\hline\hline
			$1\times 4\times 8$ in 2 ways & 206 (172)\\
			$1\times 4\times 8$ in 3 ways & 0 (0) \\
			$1\times 4\times 8$ in 4 ways & 1 (1) \\\hline
		\end{tabular}
	\end{center}
\end{table}

\clearpage
\section{Concluding Remarks}
In this work, we formulate the problem of finding a common development of plural boxes as a SAT problem, 
and obtain thousands of developments that can fold into a box (or boxes) in multiple ways.
In particular, we found a development that can fold into a $ 1 \times 2 \times 8 $ box in five different ways.
So far, the value five here is the largest
number that we have found.
We hope that analyzing developments that we have obtained would help to find a development
having a larger number of box foldings.

Our initial motivation for this work was to find a common development of three incongruent boxes whose area is
smaller than 532, which is the current known smallest \cite{SU13}. 
However, despite considerable efforts, we have not succeeded in finding such a development.
Thus, determining the smallest such development might be a challenging task.


\begin{thebibliography}{9}
\bibitem{Uehara2011}
Zachary Abel, Erik Demaine, Martin Demaine, Hiroaki Matsui, G\"{u}nter Rote, Ryuhei Uehara, Common Developments of Several Different Orthogonal Boxes, Proc. of the 23rd Canadian Conf. on Comput. Geometry (CCCG 2011), pp. 77--82 (2011).

\bibitem{Bie17}
Armin Biere,
CaDiCal, Lingeling, Plingeling, Treengeling and YalSAT Entering the SAT Competition 2017,
Proc. of SAT Competition 2017, pp. 14--15 (2017).

\bibitem{DO07}
Eric D.~Demaine and Joseph O'Rourke,
Geometric Folding Algorithms: Linkages, Origami, Polyhedra.
Cambridge University Press (2007)

\bibitem{Uehara2019}
    Koichi Mizunashi, Takashi Horiyama, Ryuhei Uehara, Efficient Algorithm for Box Folding, Proc. of the 13th Int. Conf. and Workshops on Algorithms and Computation (WALCOM 2019), pp. 277--288 (2019).
    
\bibitem{SATCompetition}
SAT Competition 2017, \url{<https://baldur.iti.kit.edu/sat-competition-2017/>} (retrieved 2020.3.27)

\bibitem{SU13}
Toshihiro Shirakawa, Ryuhei Uehara,
Common Developments of Three Incongruent Orthogonal Boxes, 
Int. J. Comput. Geometry Appl. 23(1), pp. 65--71 (2013)

\bibitem{Ueh14}
Ryuhei Uehara,
A Survey and Recent Results about Common Developments of Two or More Boxes, 
Proc. of the 6th Int. Meeting Origami in Sci., Math. and Educ. (OSME 2014), pp. 77--84 (2014)

\bibitem{XH17}
Dawei Xu, Takashi Horiyama, Toshihiro Shirakawa and Ryuhei Uehara,
Common Developments of Three Incongruent Boxes of Area 30,
Computational Geometry, 64, pp. 1--12 (2017)

\end{thebibliography}
\end{document}